\documentclass[twocolumn,prd,nofootinbib,showpacs]{revtex4}
\renewcommand{\vec}[1]{\mathbf{#1}}

\usepackage{graphicx}

\newcommand{\beq}{\begin{equation}}
\newcommand{\eeq}{\end{equation}}
\newcommand{\beqa}{\begin{eqnarray}}
\newcommand{\eeqa}{\end{eqnarray}}
\begin{document}
\title{Contamination cannot explain the lack of large-scale power
in \\ the cosmic microwave background radiation}
\author{Emory F. Bunn}
\email{ebunn@richmond.edu}
\author{Austin Bourdon}
\affiliation{Physics Department, University of Richmond, Richmond, VA  23173}

\begin{abstract}
Several anomalies appear to be present in the large-angle cosmic
microwave background (CMB) anisotropy maps of WMAP.  One of 
these
is a lack of large-scale power.  Because the data
otherwise match standard models extremely well, it is natural to
consider perturbations of the standard model as possible explanations.
We show that, as long as the source of the perturbation is
statistically independent of the source of the primary CMB anisotropy,
no such model can explain this large-scale power deficit.
On the contrary, any such perturbation always {\it reduces} the
probability of obtaining any given low value of large-scale power.
We rigorously prove this result when
the lack of large-scale power is quantified
with a quadratic statistic, such as the quadrupole moment.  When
a statistic based on 
the integrated square of the correlation
function is used instead, we present strong numerical evidence 
in support of the result.
The result applies to 
models in which the geometry of spacetime is perturbed (e.g., 
an ellipsoidal Universe) as well as explanations
involving local contaminants, undiagnosed foregrounds, or systematic errors.
Because the large-scale power deficit is arguably the most significant
of the observed anomalies, explanations that worsen this discrepancy
should be regarded with great skepticism, even if they help in explaining
other anomalies such as multipole alignments.
\end{abstract}
\maketitle

\section{Introduction}

Observations of cosmic microwave background (CMB) anisotropy,
particularly the data from WMAP
\cite{wmap1yr1,wmap1yr2,wmap1,wmap5yrbasic}, have revolutionized
cosmology.  These observations are a major contributor to the
emergence of a cosmological ``standard model'' of a Universe dominated
by dark energy and cold dark matter, with a nearly scale-invariant
spectrum of Gaussian adiabatic perturbations
\cite{wmap5yrparams,wmap5yrinterp}.  The overall consistency of the
CMB data with this model is quite remarkable, but there appear to be
some anomalies on the largest angular scales,
such as
a lack of
large-scale power \cite{wmap1yr2,copi2,dOCTZH}, alignment of low-order multipoles
\cite{schwarz,copi1,dOCTZH,hajian}, and hemispheric asymmetries
\cite{eriksen2004,freeman}.

The significance of and explanations for these puzzles
are hotly debated.  In particular, it is difficult to know how to interpret
a posteriori statistical significances: when a statistic is invented
to quantify an anomaly that has already been noticed, the low $p$-values
for that statistic cannot be taken at face value.
Nonetheless, the number and nature of the anomalies (in particular, the
fact that several seem to pick out the same directions on the sky)
seem to suggest that there may be something to explain in the data.
In this paper, we will tentatively assume that there is a need for an 
explanation and consider what that explanation might be.

Since the standard model is in general highly consistent with the
CMB and a wide variety of other observations, it is natural
to look for explanations of these puzzles that consist of
perturbations added onto the standard model.  Such explanations can be based 
on nonstandard cosmologies, such as
ellipsoidal models \cite{ellipsoid}, large-scale magnetic fields
\cite{barrow}, and 
theories based
on Bianchi VIIh spacetimes with rotation and shear \cite{ghosh,bianchi}.
They can also involve phenomena on much smaller scales (e.g., \cite{inoue,
inoue2,abramo,cooray}),
perhaps even within the Solar System \cite{frisch}.
Any uniagnosed foreground contaminant 
would fall into the class
of explanations we consider, as would many systematic errors.

All of these models can be described
by assuming that
the observed CMB sky is the sum
of two terms: 
\beq
T_{\rm obs}({\bf r})=T_0({\bf r})+T_c({\bf r}),
\eeq
where $T_0$ is a Gaussian CMB sky with a power spectrum given
by the standard model and $T_c$ is a contaminant.
The contaminant can 
be a fixed function of sky position $\bf r$ or a realization
of a random process.  In the latter case, 
we assume nothing about the statistics of this process
except that it is independent of the Gaussian random process that produced
$T_0$.
We wish to consider the possibility that such a model can explain
some or all of the large-angle anomalies.  

In this article, we will present strong evidence that on the contrary
all such models actually exacerbate one of the anomalies, namely
the observed lack of power in the large-angular-scale CMB anisotropy.
This anomaly is formally highly statistically significant, and as we will
argue below it is one for which the problems of a posteriori statistics
are not particularly severe.  It is therefore arguably the most in need of explanation
of all of the large-angle CMB puzzles.  We conclude, therefore, that
this entire category of possible explanations should be regarded
with great skepticism.  In particular, the absence of large-scale power
in the WMAP data is in fact a strong argument {\it against} 
the existence of undiagnosed foreground contamination,  as well as systematic
errors that would produce an additive contaminant to the observed
sky maps.

\begin{figure}
\includegraphics[width=3.5in]{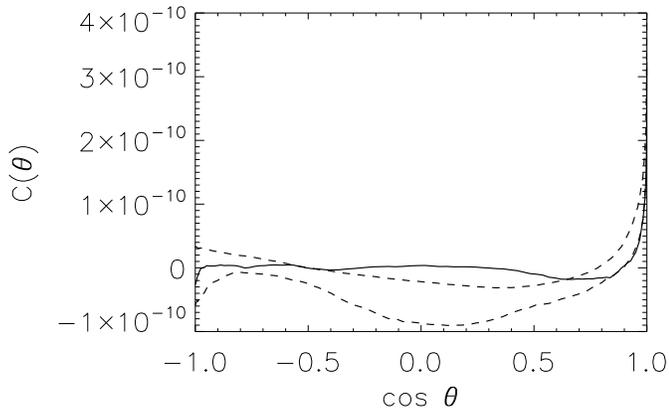}
\caption{The two-point correlation function for the WMAP data.  The
solid curve shows the correlation function for the three-year WMAP
internal linear combination (ILC) data at resolution
$N_{\rm side}=32$, computed with the SpICE 
software \cite{spice}.  The dashed curves show the 95\% confidence
range in a set of simulations.  The simulations virtually never produce
correlation functions that are as close to zero at large angles as
the real data.}
\label{fig:corrfunc}
\end{figure}

We can quantify the lack of power in the large-angle CMB by
considering either the power in low-order multipoles (especially
the quadrupole) or a statistic based on
the two-point angular correlation function (see Fig.~\ref{fig:corrfunc}).
In either case, the $p$-values (that is, the probabilities of getting
as low a value of the chosen statistic as the one in the actual data)
are low; in fact, 
for some choices of statistic, they are less than 0.1\% \cite{copi2} (but
see \cite{efstathiou2} for a constrasting analysis).  
By definition, for an alternative theory to explain this anomaly, it would have
to generate larger $p$-values.  We will show in this paper that all proposed
models of the form described above in
fact {\it reduce} the $p$-values based on these statistics.
Therefore, although such models might alleviate some of the other
large-angle anomalies, they worsen this one.

At one level, this is not surprising.  For the models considered here,
in which the observations are the sum of two statistically independent
terms, the observed power spectrum is simply the sum of the 
standard-model spectrum and the spectrum of the contaminant.  Addition
of the contaminant therefore biases all multipoles up, including
the quadrupole.  This is merely a statement about mean-square values,
however, and does not tell us about the probability distribution of the
multipoles.  
It is logically possible that a (non-Gaussian) contaminant, even as
it biases the mean-square quadrupole up, widens the probability distribution
for the quadrupole
in such a way as to enhance the probability of getting low values.
Indeed, any proposal to explain the lack of large-scale power
through 
a perturbation to the standard model must be proposing such
an effect, since this is what it would mean to ``explain'' the discrepancy.  

For example, suggestions have been made that
the low quadrupole might be explained by an extended local
foreground \cite{abramo}, by dust-filled local voids\footnote{A suggestion
is made in the cited 
work that the hypothesis that the contaminant is uncorrelated
with the primary signal may not apply.  If this is true, then the 
arguments in the present paper would not apply to this model.  It is
not clear to us that a strong correlation of the proposed form exists
in the model under consideration, and as far as we know no detailed
calculation of this effect has been performed.} \cite{inoue,inoue2},
or by an ``ellipsoidal'' universe that
expands at different rates in different directions \cite{ellipsoid}.
Each such explanation assumes that
a chance anticorrelation between the contaminant and the intrinsic
CMB anisotropy has occurred.  In order for this to count as an explanation,
however, such an anticorrelation must be sufficiently probable
that it raises the probability of finding the observed lack of power.  
Although this is a logical possibility, we will argue below that it in
fact never occurs, whether the lack of power is quantified via
the quadrupole moment or the correlation function.  
For some specific cases, such as the quadrupole moment in an 
ellipsoidal universe,
previous work \cite{ellipsoidiswrong} 
has already established this; in this paper we prove it in general.
In summary,
such models cannot explain the lack of
large-scale power, and in fact always ``anti-explain'' it by reducing
the already-low probability.  

Section \ref{sec:quadratic} proves this general result in the case where
the lack of power is quantified via a quadratic estimator such as
the mean-square quadrupole moment.  Section \ref{sec:correlation} presents
strong numerical evidence that the result is also true
in the case of a statistic based on the two-point correlation function.
Section \ref{sec:discussion} contains a brief discussion of the results,
and an appendix proves a key mathematical result needed in section 
\ref{sec:quadratic}.

\begin{figure*}[t]
\includegraphics[width=3.5in]{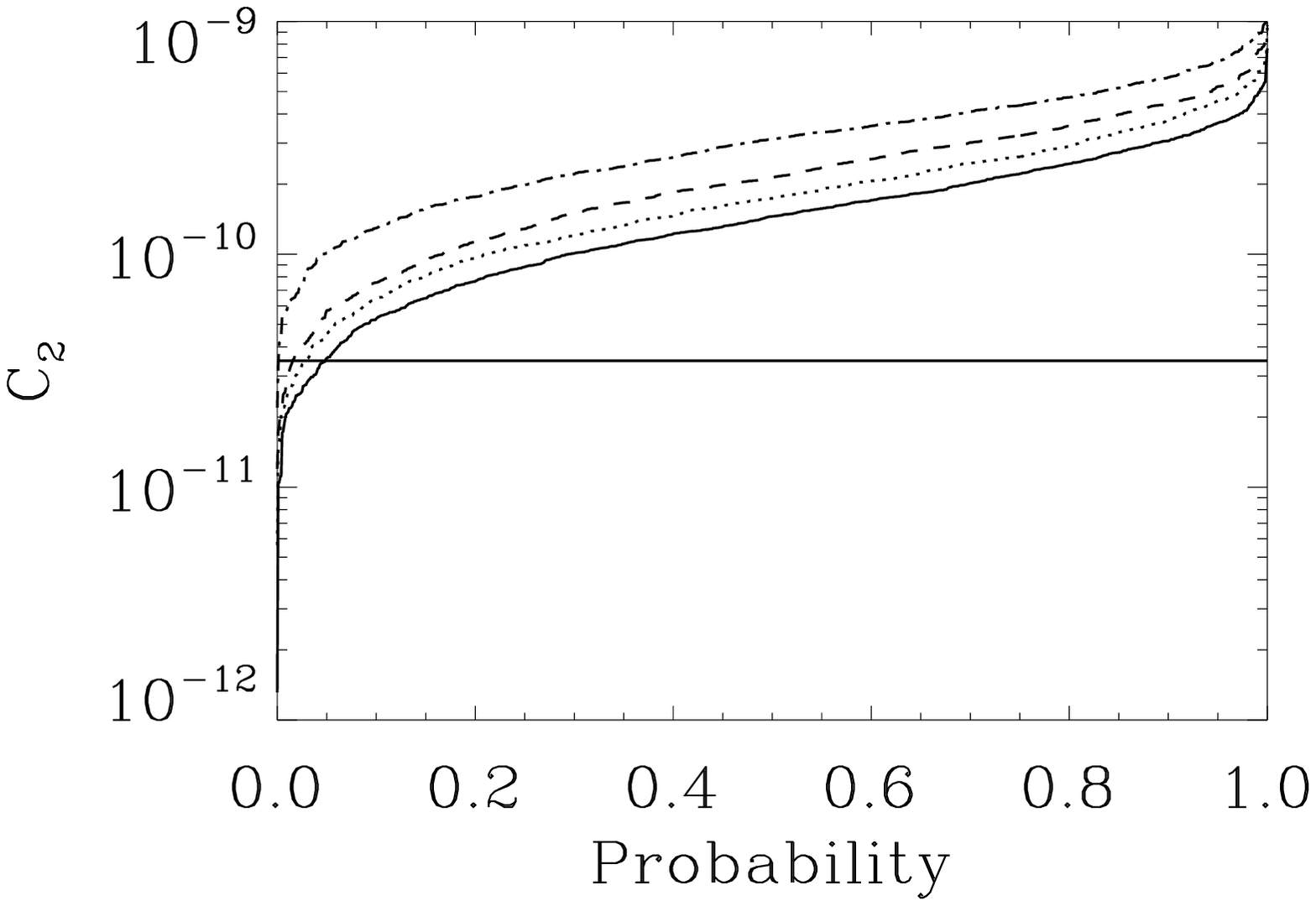}
\includegraphics[width=3.5in]{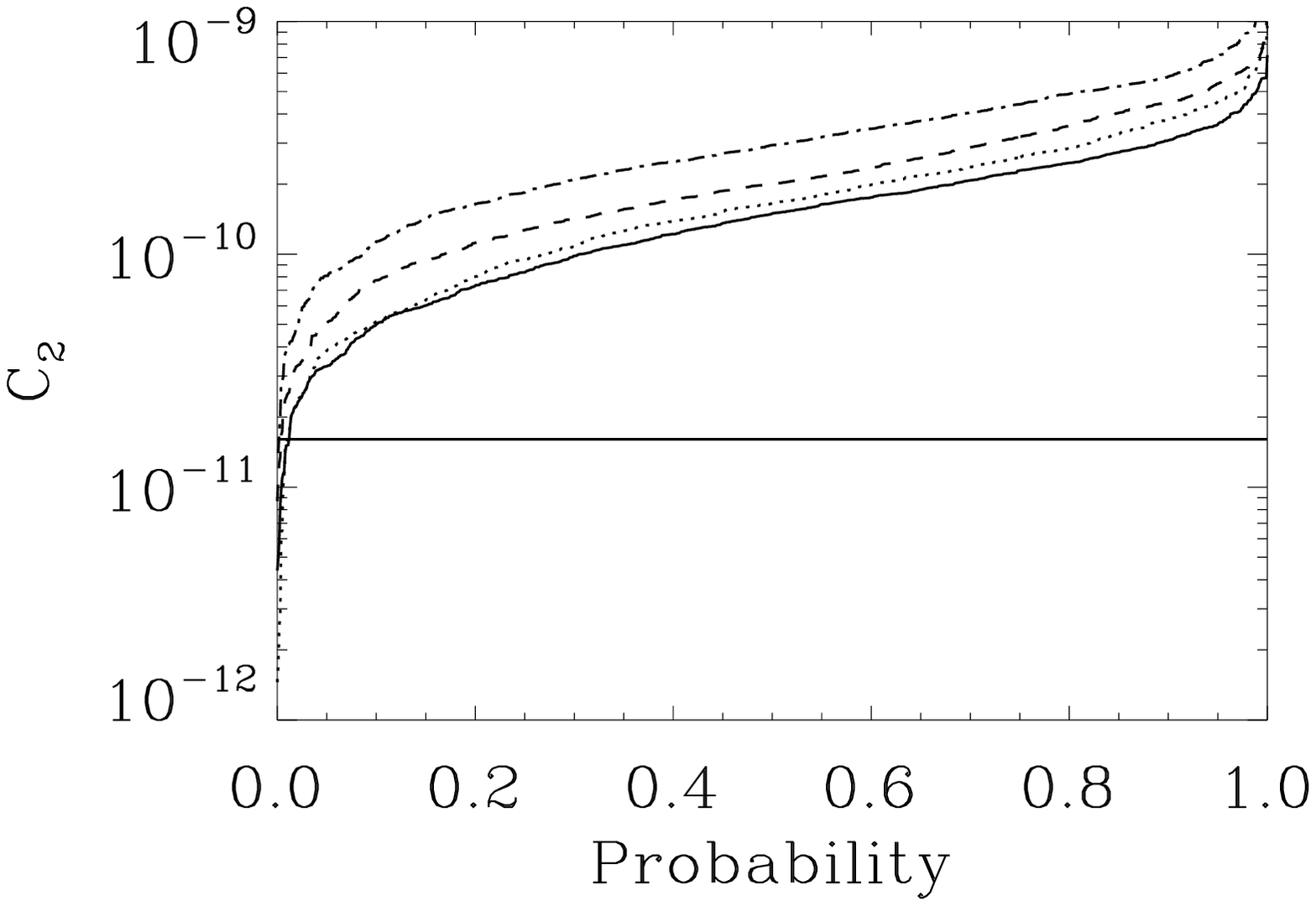}
\caption{
Cumulative probability distributions for the quadrupole
power $C_2$ in an ellipsoidal Universe, in dimensionless $\Delta T/T$
units.  In the left panel,
data from the entire sky were used, and in the right panel
the WMAP Kp0 cut was applied.  The solid curve shows
a standard LCDM model with no eccentricity.  From
bottom to top, the other three curves correspond to eccentricities
$5\times 10^{-3},6.2\times 10^{-3},7.4\times 10^{-3}$.  
The horizontal line shows the quadrupole found in the actual data.
}
\label{fig:C2}
\end{figure*}
\section{Quadratic Power Estimators}\label{sec:quadratic}
As noted above, the observed lack of large-scale power in the CMB
can be quantified in different ways.  The simplest, going all the way
back to the COBE observations \cite{gouldquad,cobe}, is to compute an estimator
of the quadrupole power $C_2=\langle|a_{2m}|^2\rangle$, where $a_{lm}$ is
a coefficient in a spherical harmonic expansion.  Quadrupole
estimators applied to 
the WMAP data are lower than theoretical predictions, although
due to the large cosmic variance, the significance of this anomaly
is only $\sim 5\%$ \cite{dOCTZH}, which is weaker than the correlation
function statistic described in the next section.  Nonetheless, because
the quadrupole is one of the simplest and most natural ways to quantify
large-scale power, we consider it in detail in this section.
In particular, we
will demonstrate that any statistically independent 
contaminant exacerbates the problem of an
anomlaously low quadrupole.

The quadrupole power is a positive definite quadratic function $q^2$ of the
data.  As noted in the previous section, 
a contaminant always causes an upward
bias in the expectation value of such a statistic; to be precise, the
expectation value is $\langle q^2\rangle= \langle q_0^2\rangle+\langle
q_c^2\rangle$, where the two terms on the right are the expectation
values due to the to contributors $T_0,T_c$.  As noted 
in the previous section, however, this statement is not
sufficient to justify the claim that adding a contaminant always
exacerbates the problem of a low quadrupole.  We need 
to show that the probability of getting a low quadrupole
is always reduced by adding a contaminant -- that is, 
for any given value $\hat q^2$,
the probability that the observed
value is less than $\hat q^2$ is always lower with a contaminant
than without.

Let the vector $\vec y$ represent a list of data points
that we will use to estimate the large-angle power in the 
CMB, for example, 
the 
pixelized temperature values in the WMAP data.  Let $q^2$
be a positive definite quadratic function of the data
(possibly with some noise bias removed):
\beq
q^2(\vec y)=\vec y \cdot{\bf A}\cdot\vec y - b.
\label{eq:quadratic}
\eeq
Here ${\bf A}$ is a symmetric nonnegative definite matrix, and the
noise bias $b$ is 
a constant.  

We want to compare the null hypothesis, that $\vec y$ contains
only intrinsic CMB anisotropy and noise, with the hypothesis
that there is an additional statistically independent contaminant.
%\begin{enumerate}
%\item The data vector $\vec y$ contains only intrinsic CMB anisotropy
%and noise.
%\item The data contains an extra contaminant that is statistically
%independent of the intrinsic anisotropy.
%\end{enumerate}
We can express these possibilities by writing
\beq 
\vec y=\vec x+\vec c, 
\eeq 
where $\vec x$ is the
``uncontaminated'' data (including noise)
and $\vec c$ represents a hypothetical contaminant.  
We assume that $\vec x$
is drawn from a multivariate Gaussian distribution:
\beq
f_x(\vec x)\propto\exp\left(-{1\over 2}\vec x\cdot {\bf M}\cdot\vec x
\right)
\eeq
for some symmetric positive definite matrix ${\bf M}$.  For the null
hypothesis, we set $\vec c=0$.
When
considering contamination, we assume $\vec c$ is a random variable
with some probability density $f_c$.  (This formulation includes
the possibility that $\vec c$ is a fixed contaminant -- that is,
$f_c$ is allowed to be a delta function.)
No assumption is made about $f_c$ other than independence
of $\vec x$ and $\vec c$, which means that the joint probability
density factors:
\beq
f(\vec x,\vec c)=f_x(\vec x)f_c(\vec c).
\eeq

Let $\hat{\vec y}$ be the data actually measured,
and let $\hat{q}^2=q^2(\hat{\vec y})$ 
stand for the power estimate obtained from it.
Let $P_{\vec{c}}$ stand for the probability of getting
a value of $q^2$ as low as the true value, assuming a fixed
value for the contaminant $\vec c$:
\beq
P_{\vec c}=\mbox{Pr}
[q^2(\vec y)<\hat{q}^2\ | \ \vec{c}]=\int_{(\vec x+\vec c)\in V}
d\vec x\,f_x(\vec x),
\label{eq:Pc}
\eeq
where the volume $V$ is the ellipsoid consisting of all $\vec y$
with $\vec y\cdot{\bf A}\cdot\vec y<\hat{q}^2+b$.

Note
that $P_{\vec c}$ is an integral
over an ellipsoid
centered at $\vec x=-\vec c$.  Since
the integrand peaks at the origin, we would expect $P_{\vec{c}}$ 
to be maximized
when the ellipsoid's center is placed at the origin.
To be specific, we expect that
\beq
P_{\vec c}\le P_0.
\label{eq:inequality}
\eeq
This expectation is indeed correct; a proof
of it may be found in the Appendix.

\begin{figure*}[t]
\includegraphics[width=3.5in]{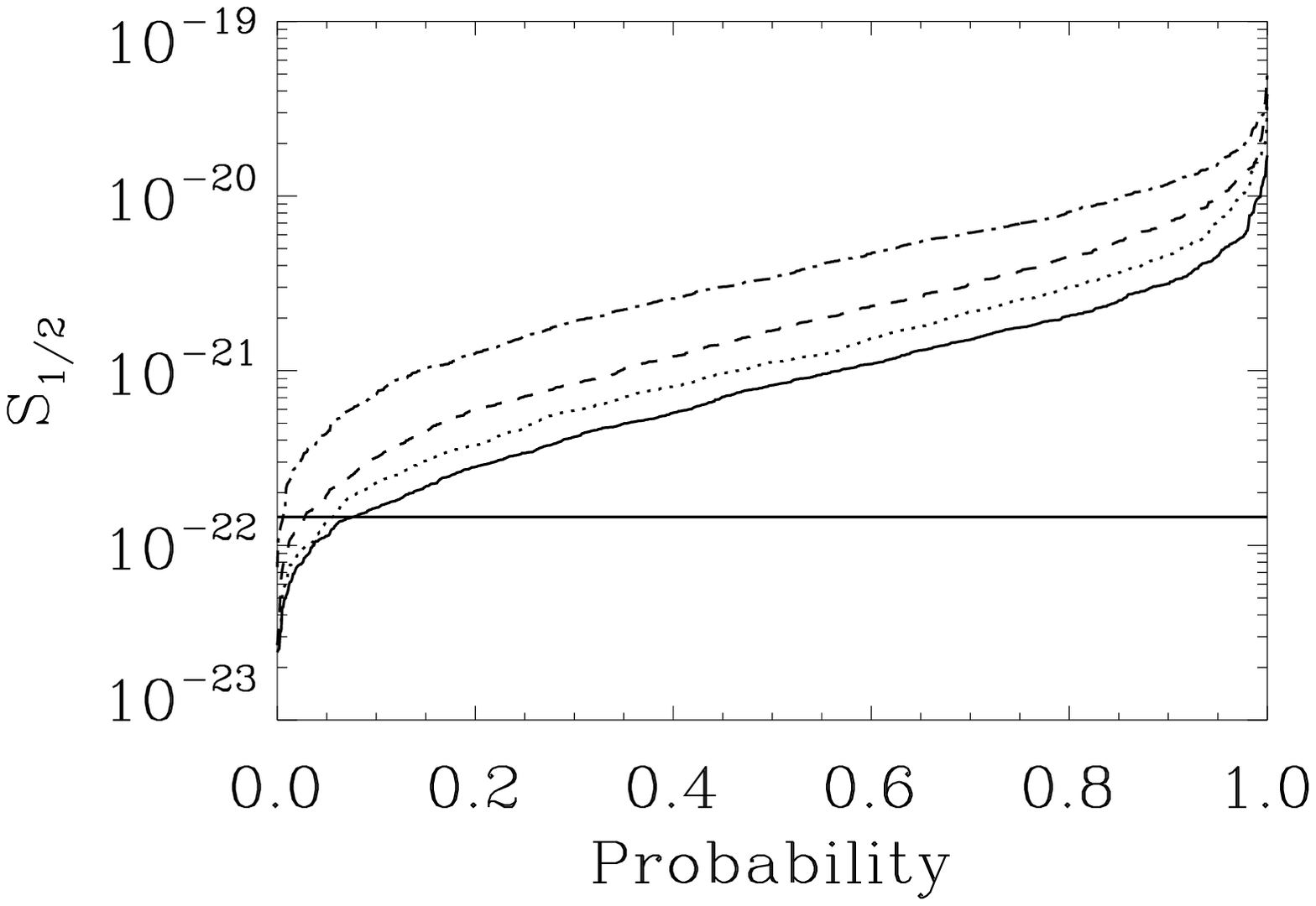}
\includegraphics[width=3.5in]{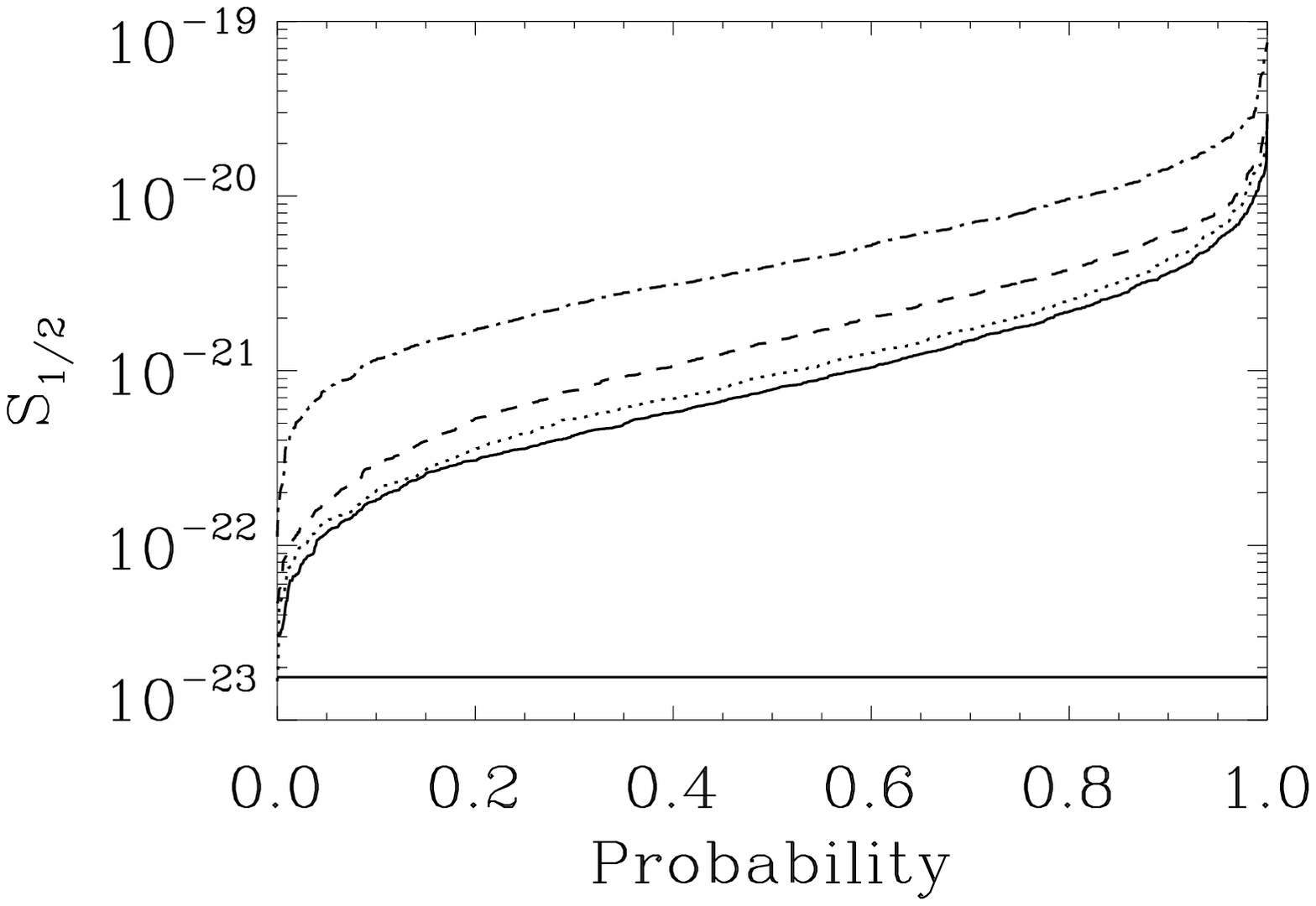}
\caption{Cumulative probability distributions for the statistic
$S_{1/2}$.  As in Figure \ref{fig:C2}, the left panel is
for full-sky data, while the right panel is for the Kp0 cut.  Curves
are as in the previous figure.}
\label{fig:corr1}
\end{figure*}

This means that adding any fixed contaminant $\vec c$ always
reduces the probability of getting a low $q^2$.  As a consequence,
even if $\vec c$ is not fixed but is generated by some random
process, the probability is still lower than in the case $\vec c=0$.
Formally, we can write
\beq
\mbox{Pr}[q^2(\vec y)<\hat{q}^2]=
\int d\vec c\,\mbox{Pr}[q^2(\vec y)<\hat{q}^2\ |\ \vec{c}]f_c(\vec c).
\eeq
%making use of the hypothesis that $\vec c$ is independent of $\vec x$.
Using inequality (\ref{eq:inequality}),
\beq
\mbox{Pr}[q^2(\vec y)<\hat{q}^2]\le P_0\int f_c(\vec c)\,dc=P_0.
\label{eq:inequality2}
\eeq
%\beq
%P_c=P[q(\vec x+\vec c)<q_0]=
%\int_{(\vec x+\vec c)\in V} 
%d\vec x\,d\vec c\,f_x(\vec x)f_c(\vec c)
%=\int d\vec c\,f_c(\vec c)I(\vec c),
%\label{eq:pc}
%\eeq
%where
%\beq
%I(\vec c)=\int_{(\vec x+\vec c)\in V}d\vec x\,f_x(\vec x)
%\label{eq:Idef}
%\eeq
%
%
%Comparison of equations (\ref{eq:P0}) and (\ref{eq:Idef}) shows
%that $I(0)=P_0$.  Using this inequality in equation (\ref{eq:pc}),
%we find that
%\beq
%P_c\le 
%P_0\int d\vec c\,f_c(\vec c)
%=P_0,
%\eeq
%since the probability density $f_c$ integrates
%to 1.

This inequality
is the central result of this section.  It means that,
if the power is anomalously
low under the assumption of no contamination, then introducing
a contaminant can only make the problem worse.  

\begin{figure*}[t]
\includegraphics[width=3in]{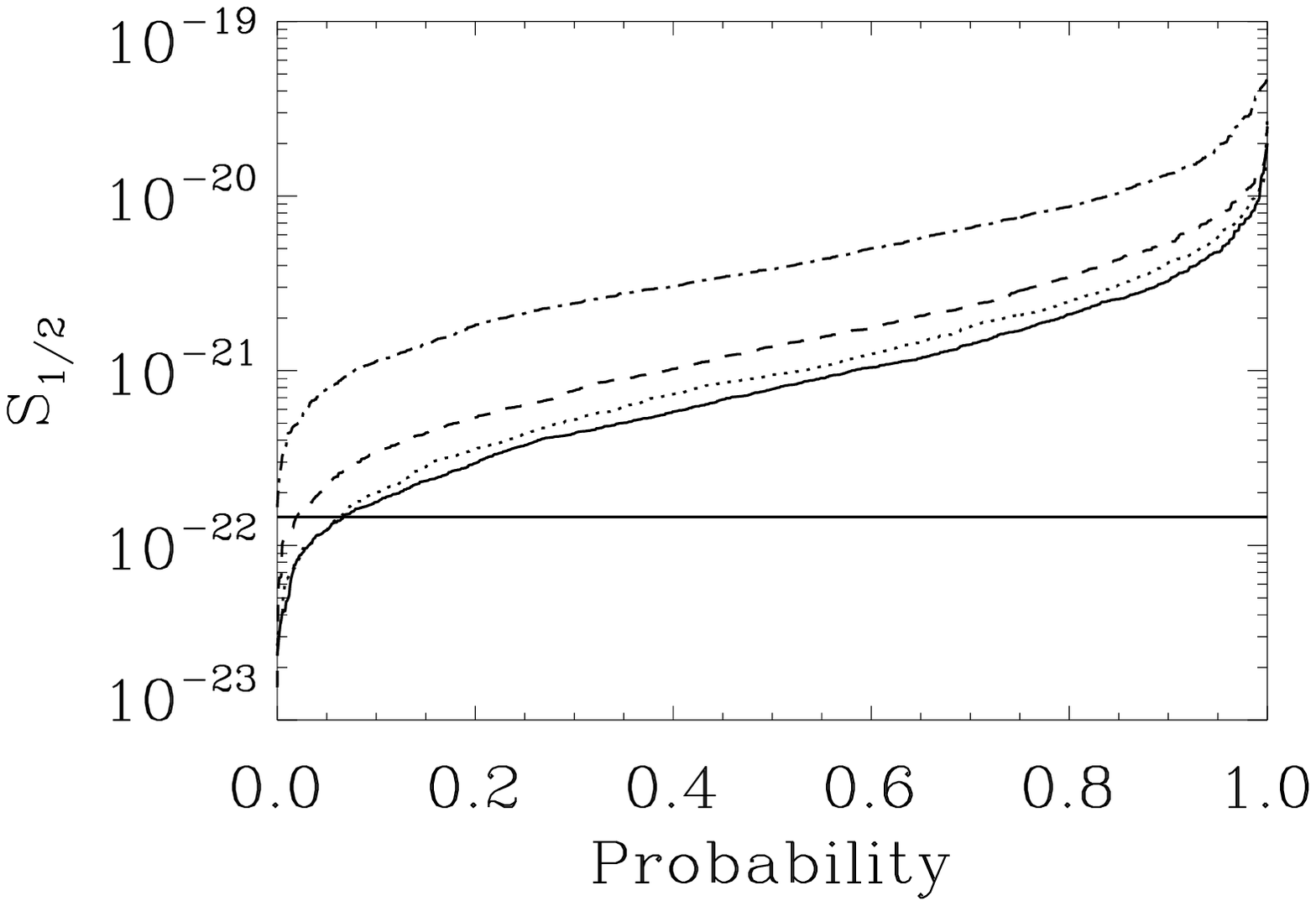}
\includegraphics[width=3in]{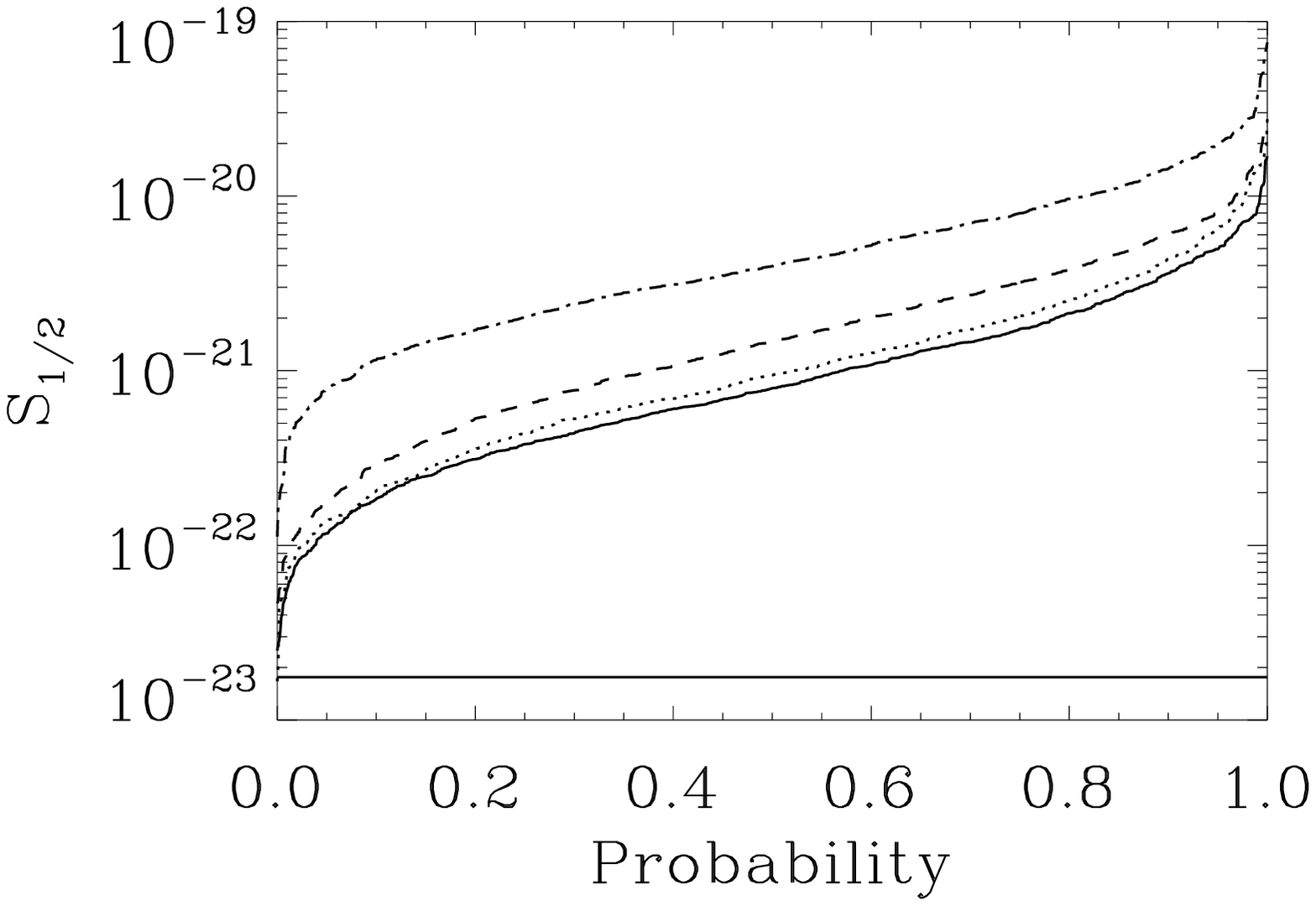}
\caption{Cumulative probability distributions
for the statistic $S_{1/2}$ for Bianchi VIIh (rotating Universe) models.
The values of the shear are 0 (solid), $2.4\times 10^{-10}$ (dotted),
$5\times 10^{-10}$ (dashed), and $1\times 10^{-9}$ (dot-dashed).
As in the previous
figures, no-cut probabilities are shown on the left, and Kp0-cut
probabilities are shown on the right.}
\label{fig:corr2}
\end{figure*}

Figure \ref{fig:C2} illustrates this conclusion for the
case of an ellipsoidal Universe.  The figure shows
the cumulative probability distribution of the quadrupole
power $C_2$, based on 1000 simulations of the CMB sky.
The simulations were performed using HEALPix \cite{healpix} with
$N_{\rm side}=32$.  The solid curve
shows the distribution for a Gaussian CMB with the power spectrum
given by the best-fit LCDM model from the
three-year WMAP data \cite{wmap2}.
From bottom to top, the other three curves show models with
the same power spectrum but with
eccentricities $5\times 10^{-3},6.2\times 10^{-3}, 7.4\times 10^{-3}$.
According to the analysis of ref.~\cite{ellipsoid}, eccentricities
in this range provide a better fit to the CMB quadrupole
than the standard model; however, as Gruppuso \cite{ellipsoidiswrong}
has pointed out, the calculations in ref.~\cite{ellipsoid} do not
properly account for all possible relative orientations of the 
ellipticity axis and the intrinsic CMB anisotropy and hence overestimate
the goodness of fit of the ellipsoidal models.
The horizontal line indicates the value found in the actual WMAP data
(specifically, the three-year internal linear combination data, downgraded
to $N_{\rm side}=32$).
The curves in the left panel were computed using the entire sky,
while the right curves were computed using the WMAP Kp0 cut \cite{wmap1}.

The figure illustrates that the probability of getting
a quadrupole value below any given cutoff strictly decreases
as the size of the perturbation increases.  As predicted by inequality 
(\ref{eq:inequality2}), the way to get the highest probability is to
have no perturbation at all.
In particular,
for the no-cut data, the probability of getting a value
as small as the actual data is $\sim 5\%$ in the standard model
and dropts to $\sim 3\%,1.5\%,0.2\%$ as the ellipticity increases.
When the Kp0 mask is applied, the probabilities are lower in all cases
than in the full-sky case, but the same decrease in probability is observed.
These conclusions are consistent
with those of ref.~\cite{ellipsoidiswrong}, but we have established
the conclusion for a much broader category of theories, not just
this specific case.

\section{Correlation Function}
\label{sec:correlation}

The low quadrupole does not have particularly high statistical
significance, largely because of the high level of cosmic variance
in the quadrupole.  The two-point angular
correlation function provides a much more signficant indication that there
is an anomalous lack of large-scale power in the WMAP data.
In
particular, the integrated square of the correlation function,
\beq
S_{1/2}=\int_{-1}^{1/2}[C(\theta)]^2\,d\cos\theta,
\eeq
which was first introduced in the analysis of the 1-year WMAP
data \cite{wmap1yrparams},
is extremely low in the WMAP data in comparison with theoretical
estimates, with $p$-value of order 0.1\% \cite{copi2}.
Here $C(\theta)$ is the two-point correlation function, that is,
the average of all pairs of pixels with angular separation $\theta$.
We wish to examine whether adding a perturbation to the standard
model can solve this problem (that is, raise the probability of
getting the observed low value of $S_{1/2}$).

Since the statistic $S_{1/2}$ is quartic, not quadratic, in
the data, 
the 
argument of the previous section does not apply to it.
However, it is extremely plausible to suppose that a similar conclusion
should hold, since any model with a high probability of producing low 
values of this statistic would presumably produce low values of the
low-order multipoles, and since any contaminant reduces the probability 
of such low multipoles.

We can of course test this conjecture numerically for any
particular model.  For example, Figure \ref{fig:corr1}
shows the results of simulations precisely like those shown in Figure 
\ref{fig:C2}, but with the statistic $S_{1/2}$ used in place of the
quadrupole.  The SpICE software \cite{spice} was used to compute
the correlation functions.
Figure \ref{fig:corr2} shows the results of similar calculations,
for the case of a model in which the spacetime geometry
is that of a rotating Bianchi VIIh model \cite{bianchi}.
We have also performed computations for models in which the
contaminant consists of circular
hot and cold spots of varying amplitudes and radii, to simulate the
effects of local voids or similar features.  
In all of these cases, the addition of a
contaminant does not solve the problem of the
lack of large scale power; in fact, it worsens it.

Rather than examining theories one at a time, it would clearly
be better to have a general argument that applied to a broad class
of theories.  In the rest of this section, we provide such an argument.

Suppose that the value of $S_{1/2}$ for the actual data
is $\hat{S}_{1/2}$.  Let $V$ be the volume in the data space that
yields values of the statistic this low:
\beq
V=\{\vec y\ | \ S_{1/2}(\vec y)<\hat S_{1/2}\}.
\eeq
Then, assuming a contaminant given by a fixed
vector $\vec c$, the probability
of getting such a low value of the statistic is
\beq
P(\vec c)=\int_{(\vec x+\vec c)\in V} f_x(\vec x)\,d\vec x
=\int_{\vec y\in V}f_x(\vec y-\vec c)\,d\vec y.
\eeq
We want to know whether there are any vectors $\vec c$ such 
that $P(\vec c)>P(0)$, or in other words whether $P$ has a global
maximum at $\vec c=0$.  It is straightforward to check that
$\nabla P(0)=0$.  We next consider whether the 
point $\vec c=0$ is a maximum, a minimum, or a saddle point.
If we find that it is a maximum, then the addition of any
small contaminant worsens the problem we are trying to solve.

To answer this question, we naturally consider the matrix 
of second derivatives:
\beq
H_{jk}=-\left.{\partial P\over\partial c_j\partial c_k}\right|_{\vec c=0}.
\eeq
Then $P$ has a local maximum at the origin
if and only if ${\bf H}$ is positive definite.  Moreover, if ${\bf H}$
is not positive definite, then the eigenvectors corresponding to negative
eigenvalues yield the directions in data space
(i.e., particular forms for the contaminant $\vec c$) 
that alleviate the problem of low $S_{1/2}$.

To calculate these derivatives,
it is convenient to transform the data to a basis that diagonalizes
the covariance matrix in the Gaussian probability density $f_x$.
The most natural way to accomplish this is to work in the 
spherical harmonic basis, in which case each data point is
a coefficient $a_{lm}$.  We can normalize each data point
according to the power spectrum, setting $x_j=a_{lm}/C_l^{1/2}$,
where the index $j$ runs over all pairs $lm$.
In this case
the covariance matrix is simply the identity matrix,
and the second derivative matrix elements can be written
\beq
H_{jk}=-\int_V d\vec x\,f_x(\vec x)(x_jx_k-
\delta_{jk}).
\eeq

This integral over the many-dimensional data space can 
most easily be be estimated by Monte Carlo integration.  To be
specific, we draw
vectors $\vec x$ from the appropriate Gaussian distribution,
calculate the corresponding values of $S_{1/2}$, and use the results to
throw away all vectors that lie outside of $V$.
For all the rest, we average together the quantities $(x_jx_k-\delta_{jk})$.

In performing this Monte Carlo integration, we consider
HealPIX maps with $N_{\rm side}=32$ and the same power spectrum
as in the previous section.  
We apply Gaussian smoothing
with a 20$^\circ$ FWHM beam to the simulated maps.  This amount
of smoothing results in significant suppression (by more than $e^{-1}$) 
of spherical harmonics coefficients $l\ge 10$.  Without significant 
smoothing, fluctuations in high-$l$ modes cause significant error in
the Monte Carlo calculation even at low $l$.  The problem of anomalously
low $S_{1/2}$ persists at about the same significance ($p$-values $\simeq 
0.1\%$)
even with such
smoothing, so this smoothing does not weaken our ability to draw conclusions
about possible explanations for the anomaly.

\begin{figure}
\includegraphics[width=3.5in]{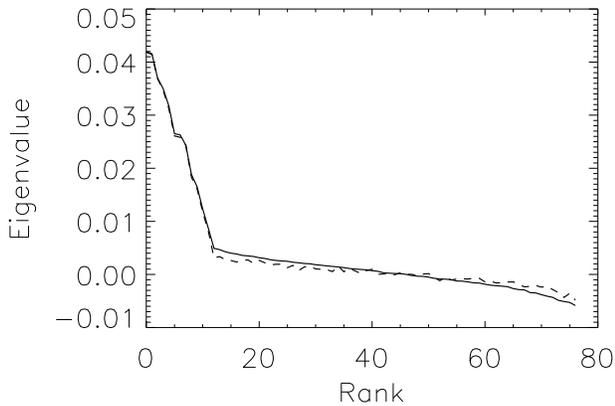}
\caption{Eigenvalues of the second derivative matrix ${\bf H}$.  Two 
independent calculations of the matrix were performed, each based
on 80\,000 simulations using the Kp0 cut.  The matrices were truncated
to include only multipoles $l=2$ through 8.  The solid curve shows the
eigenvalues computed from one matrix, sorted from
largest to smallest.  The dashed curve shows
the quantities $\vec v\cdot{\bf H}\cdot\vec v$, where $\vec v$ are the
eigenvectors computed from the first matrix and ${\bf H}$ is the second
matrix.  The difference between the two curves gives an indication of
the numerical error in the Monte Carlo integration.}
\label{fig:eigval}
\end{figure}

Figure \ref{fig:eigval} shows the eigenvalues of the matrix
resulting from this Monte Carlo integration, using the Kp0 mask.
The results look similar when data from the full sky are used.
The matrix used to compute the eigenvalues was based on 80\,000
simulations lying within the volume $V$.  
Modes up to $l=8$ were used to compute the eigenvalues
shown in the figure,
although modes up to $l=64$ (far above the beam scale) were used
in the simulations.  To test the numerical stability of the results,
we used a second set of 80\,000 simulations to recompute the matrix
${\bf H}$.  We then calculated $\vec v\cdot{\bf H}\cdot \vec v$ for
each eigenvector $\vec v$.  The results are shown in the dashed curve.
In the absence of numerical error, the two curves would be identical.

\begin{figure}
\includegraphics[width=2in,angle=90]{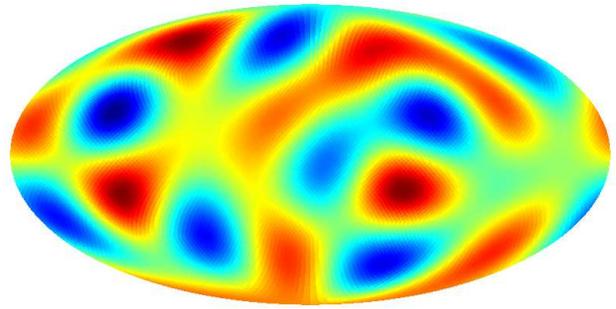}
\caption{The sky pattern corresponding to the most negative
eigenvalue of ${\bf H}$.}
\label{fig:testpattern}
\end{figure}

Although there is some numerical error due to the Monte Carlo
integration, it appears that the matrix is not positive definite.  We
wish to examine the eigenvectors corresponding to the most negative
eigenvalues, since these describe particular contaminants that might
solve the problem of a lack of large-scale power.
Figure \ref{fig:testpattern} shows the particular pattern on the
sky corresponding to the most negative eigenvalue.
Most of the power in this contaminant is found in multipole $l=5$,
as is the power in all of the most negative eigenvectors.
To test the robustness of this pattern,
we computed the eigenvectors retaining varying numbers of modes in
the matrix ${\bf H}$, ranging from $l_{\rm max}=5$ to 15, and also
using varying subsets of the Monte Carlos to compute the matrix.
The results are quite consistent, with the most negative eigenvectors 
always having most of their power at $l=5$ and looking quite similar
to Fig.~\ref{fig:testpattern}.

The existence of these negative eigenvalues seems to contradict
our assertion that no contaminant can explain the low value of $S_{1/2}$:
modes such as the one shown in Fig.~\ref{fig:testpattern}, by construction,
raise the probability of getting a low value when added to the data.
However, when we assess the amount of improvement that these modes 
can provide, we find it to be negligible.
Consider a model in which we add a contaminant of the
form shown in Fig.~\ref{fig:testpattern} with some amplitude
$\alpha$ to the standard model.
The results of this section have shown that the probability
of getting a low $S_{1/2}$ is an increasing function of $\alpha$
at low $\alpha$.  However, because the eigenvalue is fairly small,
the increase might be expected to be slight.  Furthermore, for sufficiently
large value of $\alpha$, the probability must start to decrease again.

\begin{figure}
\includegraphics[width=3.5in]{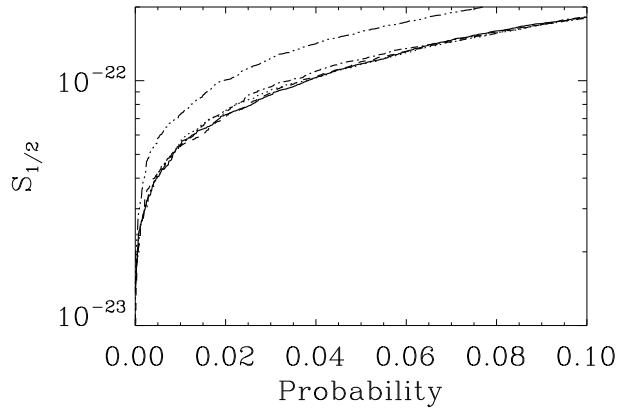}
\caption{Cumulative probability distributions for models in which
a fixed contaminant of the form shown in Fig.~\ref{fig:testpattern}
is added to the sky.  This pattern corresponds to the most negative
eigenvalue of the matrix ${\bf H}$ and so might be expected to increase
the probability of finding a low value of $S_{1/2}$.  The curves
shown in the figure are for varying amplitudes of the contaminant,
with root-mean-square pixel values of
0 (solid) 2\,$\mu$K (dotted), 4\,$\mu$K (dashed), 8\,$\mu$K (dot-dashed).
No value of the amplitude causes the probability of
getting values as low as those in the real data to increase noticeably.}
\label{fig:sprobpattern}
\end{figure}

Fig.~\ref{fig:sprobpattern} shows that this is indeed the case, and
furthermore that no choice of $\alpha$ leads to a significant increase
in the probability of getting a value as low as the actual data.  This
probability remains virtually 
unchanged at $\sim 10^{-3}$ for small $\alpha$ and
then decreases dramatically for larger $\alpha$.  Since all of the
eigenvectors corresponding to significantly negative eigenvalues of
${\bf H}$ give patterns quite similar to this one, we can conclude with
confidence that no such pattern can significantly alleviate the problem
of low $S_{1/2}$.

\section{Conclusions}\label{sec:discussion}

We have considered a broad class of cosmological models, obtained by
adding a contaminant to the standard best-fit inflation-based model.
The only assumption we have made about the contaminant is that it is
statistically independent of the cosmological signal.  We have argued
that all such models exacerbate rather than alleviating the lack of
large-scale power in the WMAP data.  We hve proven this result to be
true when the lack of power is quantified by the quadrupole moment and
have presented strong numerical evidence in support of it when
the two-point correlation function is used.  Since the latter in particular
is discrepant at a highly significant level already, any theory that
worsens this discrepancy should be regarded with great skepticism.

In addition to exotic cosmologies such as 
models with a global ellipsoidal anisotropy,
the class of models considered herein includes more
mundane possibilities such as undiagnosed foregrounds and
many systematic errors.  In particular, since several of the
observed anomalies seem to ``pick out'' the ecliptic plane
as a preferred direction, some attention has focused on 
a local foreground as a possible explanation.  The calculations
presented here argue against such models.

In any particular model with a contaminant, of course, it is possible
that a chance cancellation between the contaminant and the intrinsic
CMB anisotropy can occur, leading to the observed lack of large-scale
power.  What we have shown is that such a chance cancellation is always
unlikely, and in particular that it is always more unlikely than 
the lack of large-scale power occurring based on cosmic variance alone,
without a contaminant.

The question of how seriously to take the various large-angle CMB
anomalies, including the lack of large-scale power as well as the
various other puzzles, has been much debated \cite{efstathiou}.  In
particular, because they are all based on a posteriori
statistics ({\it i.e.}, on statistical significances calculated after
the anomalies had already been noted), the quoted significances cannot
be taken at face value.  Arguably, however, the large-scale power
deficit suffers less from this problem of a posteriori statistics.
After all, for virtually the entire existence of the field of CMB
anisotropy studies, the two-point correlation function has been
regarded as one of the most natural statistics to use in quantifying
the level of structure in CMB maps as a function of angular scale.
For instance, upper limits on CMB anisotropy 
in the pre-COBE era were usually presented as
limits on the correlation function.  Although the particular statistic
$S_{1/2}$ is an a posteriori invention, it merely quantifies the
mean-square level of this function, which was already regarded a priori as a
natural function to compute.  Although one can certainly dispute
the extent to which the significance of the lack of large-angle
correlations is an artifact of the particular choice of statistic
(for instance, see \cite{gaztanaga}, who do not use the $S_{1/2}$ statistic
and find less significant discrepancies), nonetheless
we believe that, of
all the observed anomalies, the large-scale power deficit is one of
the most in need of explanation.

Anomalies in a data set naturally prompt thoughts of systematic errors
or contaminants in the data.  Perhaps counterintuitively, this particular
anomaly provides a strong argument against such possibilities.  In particular,
a foreground that was not removed from the data (due to having a spectrum
indistinguishable from the CMB, for example) would fall precisely into
the category considered herein.  Note, however, that if the foreground
removal procedure itself removes part of the cosmological signal, 
the resulting error would not fall into the category considered herein.
In particular, the ILC method does project out some of the intrinsic
CMB signal and so in principle does reduce the amount of large-scale power.
This effect is calculable and has been found to be negligible, however.

There are of course a wide variety of possible explanations for the anomalies
that do not fall into the category considered here.  For example,
simply modifying
the primordial power spectrum at large scales naturally alleviates
the problem of a lack of large-scale CMB power (e.g., \cite{cline,contaldi}).
Some models with nontrivial topology 
(e.g., \cite{starobinsky,stevens,angelica,bonehead,luminet}) also
have this effect, although such models have other problems \cite{circles}.
The framework of spontaneous isotropy breaking \cite{gordon} also provides
a class of models that are not based on simply adding a perturbation
to the standard cosmology.  Models such as these (and many others) may
provide an explanation for the puzzles in the large-angle CMB.

\begin{acknowledgments}
This work was supported by National Science Foundation Award AST-0507395,
by the Research Corporation, and by the Virginia Foundation for Independent
Colleges.  We acknowledge use of the HEALPix \cite{healpix}
and SpICE \cite{spice} software.
\end{acknowledgments}

%\begin{widetext}
\appendix
\section*{Appendix: Proof of equation (\ref{eq:inequality})}
\label{sec:proof}

We first express equation (\ref{eq:Pc}) in terms of the
integration variable $\vec y=\vec x+\vec c$:
\beq
P_{\vec c}=\int_{\vec y\in V}d\vec y\,f_x(\vec y-\vec c).
\eeq
We next apply a linear coordinate transformation
that maps the ellipsoid $V$ onto the unit sphere.  To be specific,
we find a matrix ${\bf L}$ such that ${\bf A}={\bf L}\cdot{\bf L}^T$
(e.g., by Cholesky decomposition).  We define $\vec y'=({\bf L}^T\cdot\vec y)
/\sqrt{q_0^2+b}$ and $\vec c'$ similarly.  Then
\beq
P_{\vec c}\propto\int_{|\vec y'|^2\le 1}d\vec y'\,f_{x'}(\vec y'-\vec c')
\eeq
Here $f_{x'}$ is a multivariate Gaussian probability density with
a new inverse covariance matrix ${\bf M}'$, and the proportionality
constant is determined by the Jacobian of the coordinate transformation.
For convenience, we now make yet another coordinate transformation:
we apply a rotation that diagonalizes ${\bf M}'$.  The result is
\beq
P_{\vec c}\propto\int_{|\vec y''|<1}d\vec y''\exp\left(-\sum{(y_i''-c_i'')^2
\over 2\sigma_i^2}\right).
\eeq
For the remainder of this section we drop the double primes.

We now show that $P_{\vec c}$ has a maximum at $\vec c=0$.  
Differentiate the above expression for $P_{\vec c}$ with respect to $c_1$:
\beqa
{\partial P_{\vec c}\over\partial c_1}\hskip-0.07in & \propto& \hskip-0.07in
\int_{|\vec y|<1}dy\,\exp\left(-\sum{(y_i-c_i)^2
\over 2\sigma_i^2}\right){y_1-c_1\over\sigma_1^2}\\
&=&\int dy_2\cdots dy_n 
\exp\left(-\sum_{i=2}^n{(y_i-c_i)^2
\over 2\sigma_i^2}\right)
\times\nonumber\\
& &\quad
\int_{-Y_1}^{Y_1}\exp\left(-{(y_1-c_1)^2\over 2\sigma_1^2}
\right){y_1-c_1\over\sigma_1^2}dy_1,\\
\nonumber
\eeqa
where $Y_1=\left(1-\sum_{i=2}^ny_i^2\right)^{1/2}$.
Performing the $y_1$ integral yields
\beqa
{\partial P_{\vec c}\over\partial c_1}&\propto&
\int dy_2\cdots dy_n 
\exp\left(-\sum_{i=2}^n{(y_i-c_i)^2
\over 2\sigma_i^2}\right)\times\nonumber\\
&&\left(e^{-(Y_1+c_1)^2/2\sigma_1^2}-
e^{-(Y_1-c_1)^2/2\sigma_1^2}\right).
\eeqa
The integrand (and hence the integral) is strictly positive
for $c_1<0$ and negative for $c_1>0$.  That is, for any fixed
values of $c_2,\ldots,c_n$, the function $P_{\vec c}$ has its only maximum at 
$c_1=0$.  The same argument applies to each of the other $c_j$.
Hence $P_{\vec c}$ has a global maximum at $\vec c=0$.
%\end{widetext}

\bibliography{quad}

\begin{thebibliography}{41}
\expandafter\ifx\csname natexlab\endcsname\relax\def\natexlab#1{#1}\fi
\expandafter\ifx\csname bibnamefont\endcsname\relax
  \def\bibnamefont#1{#1}\fi
\expandafter\ifx\csname bibfnamefont\endcsname\relax
  \def\bibfnamefont#1{#1}\fi
\expandafter\ifx\csname citenamefont\endcsname\relax
  \def\citenamefont#1{#1}\fi
\expandafter\ifx\csname url\endcsname\relax
  \def\url#1{\texttt{#1}}\fi
\expandafter\ifx\csname urlprefix\endcsname\relax\def\urlprefix{URL }\fi
\providecommand{\bibinfo}[2]{#2}
\providecommand{\eprint}[2][]{\url{#2}}

\bibitem[{\citenamefont{{Hinshaw} et~al.}(2003)\citenamefont{{Hinshaw},
  {Spergel}, {Verde}, {Hill}, {Meyer}, {Barnes}, {Bennett}, {Halpern},
  {Jarosik}, {Kogut} et~al.}}]{wmap1yr1}
\bibinfo{author}{\bibfnamefont{G.}~\bibnamefont{{Hinshaw}}},
  \bibinfo{author}{\bibfnamefont{D.~N.} \bibnamefont{{Spergel}}},
  \bibinfo{author}{\bibfnamefont{L.}~\bibnamefont{{Verde}}},
  \bibinfo{author}{\bibfnamefont{R.~S.} \bibnamefont{{Hill}}},
  \bibinfo{author}{\bibfnamefont{S.~S.} \bibnamefont{{Meyer}}},
  \bibinfo{author}{\bibfnamefont{C.}~\bibnamefont{{Barnes}}},
  \bibinfo{author}{\bibfnamefont{C.~L.} \bibnamefont{{Bennett}}},
  \bibinfo{author}{\bibfnamefont{M.}~\bibnamefont{{Halpern}}},
  \bibinfo{author}{\bibfnamefont{N.}~\bibnamefont{{Jarosik}}},
  \bibinfo{author}{\bibfnamefont{A.}~\bibnamefont{{Kogut}}},
  \bibnamefont{et~al.}, \bibinfo{journal}{\apjs}
  \textbf{\bibinfo{volume}{148}}, \bibinfo{pages}{135} (\bibinfo{year}{2003}),
  \eprint{arXiv:astro-ph/0302217}.

\bibitem[{\citenamefont{{Bennett} et~al.}(2003)\citenamefont{{Bennett},
  {Halpern}, {Hinshaw}, {Jarosik}, {Kogut}, {Limon}, {Meyer}, {Page},
  {Spergel}, {Tucker} et~al.}}]{wmap1yr2}
\bibinfo{author}{\bibfnamefont{C.~L.} \bibnamefont{{Bennett}}},
  \bibinfo{author}{\bibfnamefont{M.}~\bibnamefont{{Halpern}}},
  \bibinfo{author}{\bibfnamefont{G.}~\bibnamefont{{Hinshaw}}},
  \bibinfo{author}{\bibfnamefont{N.}~\bibnamefont{{Jarosik}}},
  \bibinfo{author}{\bibfnamefont{A.}~\bibnamefont{{Kogut}}},
  \bibinfo{author}{\bibfnamefont{M.}~\bibnamefont{{Limon}}},
  \bibinfo{author}{\bibfnamefont{S.~S.} \bibnamefont{{Meyer}}},
  \bibinfo{author}{\bibfnamefont{L.}~\bibnamefont{{Page}}},
  \bibinfo{author}{\bibfnamefont{D.~N.} \bibnamefont{{Spergel}}},
  \bibinfo{author}{\bibfnamefont{G.~S.} \bibnamefont{{Tucker}}},
  \bibnamefont{et~al.}, \bibinfo{journal}{\apjs}
  \textbf{\bibinfo{volume}{148}}, \bibinfo{pages}{1} (\bibinfo{year}{2003}),
  \eprint{arXiv:astro-ph/0302207}.

\bibitem[{\citenamefont{{Hinshaw} et~al.}(2007)\citenamefont{{Hinshaw},
  {Nolta}, {Bennett}, {Bean}, {Dor{\'e}}, {Greason}, {Halpern}, {Hill},
  {Jarosik}, {Kogut} et~al.}}]{wmap1}
\bibinfo{author}{\bibfnamefont{G.}~\bibnamefont{{Hinshaw}}},
  \bibinfo{author}{\bibfnamefont{M.~R.} \bibnamefont{{Nolta}}},
  \bibinfo{author}{\bibfnamefont{C.~L.} \bibnamefont{{Bennett}}},
  \bibinfo{author}{\bibfnamefont{R.}~\bibnamefont{{Bean}}},
  \bibinfo{author}{\bibfnamefont{O.}~\bibnamefont{{Dor{\'e}}}},
  \bibinfo{author}{\bibfnamefont{M.~R.} \bibnamefont{{Greason}}},
  \bibinfo{author}{\bibfnamefont{M.}~\bibnamefont{{Halpern}}},
  \bibinfo{author}{\bibfnamefont{R.~S.} \bibnamefont{{Hill}}},
  \bibinfo{author}{\bibfnamefont{N.}~\bibnamefont{{Jarosik}}},
  \bibinfo{author}{\bibfnamefont{A.}~\bibnamefont{{Kogut}}},
  \bibnamefont{et~al.}, \bibinfo{journal}{\apjs}
  \textbf{\bibinfo{volume}{170}}, \bibinfo{pages}{288} (\bibinfo{year}{2007}),
  \eprint{arXiv:astro-ph/0603451}.

\bibitem[{\citenamefont{{Hinshaw} et~al.}(2008)\citenamefont{{Hinshaw},
  {Weiland}, {Hill}, {Odegard}, {Larson}, {Bennett}, {Dunkley}, {Gold},
  {Greason}, {Jarosik} et~al.}}]{wmap5yrbasic}
\bibinfo{author}{\bibfnamefont{G.}~\bibnamefont{{Hinshaw}}},
  \bibinfo{author}{\bibfnamefont{J.~L.} \bibnamefont{{Weiland}}},
  \bibinfo{author}{\bibfnamefont{R.~S.} \bibnamefont{{Hill}}},
  \bibinfo{author}{\bibfnamefont{N.}~\bibnamefont{{Odegard}}},
  \bibinfo{author}{\bibfnamefont{D.}~\bibnamefont{{Larson}}},
  \bibinfo{author}{\bibfnamefont{C.~L.} \bibnamefont{{Bennett}}},
  \bibinfo{author}{\bibfnamefont{J.}~\bibnamefont{{Dunkley}}},
  \bibinfo{author}{\bibfnamefont{B.}~\bibnamefont{{Gold}}},
  \bibinfo{author}{\bibfnamefont{M.~R.} \bibnamefont{{Greason}}},
  \bibinfo{author}{\bibfnamefont{N.}~\bibnamefont{{Jarosik}}},
  \bibnamefont{et~al.}, \bibinfo{journal}{ArXiv e-prints}
  \textbf{\bibinfo{volume}{803}} (\bibinfo{year}{2008}), \eprint{0803.0732}.

\bibitem[{\citenamefont{{Dunkley} et~al.}(2008)\citenamefont{{Dunkley},
  {Komatsu}, {Nolta}, {Spergel}, {Larson}, {Hinshaw}, {Page}, {Bennett},
  {Gold}, {Jarosik} et~al.}}]{wmap5yrparams}
\bibinfo{author}{\bibfnamefont{J.}~\bibnamefont{{Dunkley}}},
  \bibinfo{author}{\bibfnamefont{E.}~\bibnamefont{{Komatsu}}},
  \bibinfo{author}{\bibfnamefont{M.~R.} \bibnamefont{{Nolta}}},
  \bibinfo{author}{\bibfnamefont{D.~N.} \bibnamefont{{Spergel}}},
  \bibinfo{author}{\bibfnamefont{D.}~\bibnamefont{{Larson}}},
  \bibinfo{author}{\bibfnamefont{G.}~\bibnamefont{{Hinshaw}}},
  \bibinfo{author}{\bibfnamefont{L.}~\bibnamefont{{Page}}},
  \bibinfo{author}{\bibfnamefont{C.~L.} \bibnamefont{{Bennett}}},
  \bibinfo{author}{\bibfnamefont{B.}~\bibnamefont{{Gold}}},
  \bibinfo{author}{\bibfnamefont{N.}~\bibnamefont{{Jarosik}}},
  \bibnamefont{et~al.}, \bibinfo{journal}{ArXiv e-prints}
  \textbf{\bibinfo{volume}{803}} (\bibinfo{year}{2008}), \eprint{0803.0586}.

\bibitem[{\citenamefont{{Komatsu} et~al.}(2008)\citenamefont{{Komatsu},
  {Dunkley}, {Nolta}, {Bennett}, {Gold}, {Hinshaw}, {Jarosik}, {Larson},
  {Limon}, {Page} et~al.}}]{wmap5yrinterp}
\bibinfo{author}{\bibfnamefont{E.}~\bibnamefont{{Komatsu}}},
  \bibinfo{author}{\bibfnamefont{J.}~\bibnamefont{{Dunkley}}},
  \bibinfo{author}{\bibfnamefont{M.~R.} \bibnamefont{{Nolta}}},
  \bibinfo{author}{\bibfnamefont{C.~L.} \bibnamefont{{Bennett}}},
  \bibinfo{author}{\bibfnamefont{B.}~\bibnamefont{{Gold}}},
  \bibinfo{author}{\bibfnamefont{G.}~\bibnamefont{{Hinshaw}}},
  \bibinfo{author}{\bibfnamefont{N.}~\bibnamefont{{Jarosik}}},
  \bibinfo{author}{\bibfnamefont{D.}~\bibnamefont{{Larson}}},
  \bibinfo{author}{\bibfnamefont{M.}~\bibnamefont{{Limon}}},
  \bibinfo{author}{\bibfnamefont{L.}~\bibnamefont{{Page}}},
  \bibnamefont{et~al.}, \bibinfo{journal}{ArXiv e-prints}
  \textbf{\bibinfo{volume}{803}} (\bibinfo{year}{2008}), \eprint{0803.0547}.

\bibitem[{\citenamefont{{Copi} et~al.}(2007)\citenamefont{{Copi}, {Huterer},
  {Schwarz}, and {Starkman}}}]{copi2}
\bibinfo{author}{\bibfnamefont{C.~J.} \bibnamefont{{Copi}}},
  \bibinfo{author}{\bibfnamefont{D.}~\bibnamefont{{Huterer}}},
  \bibinfo{author}{\bibfnamefont{D.~J.} \bibnamefont{{Schwarz}}},
  \bibnamefont{and} \bibinfo{author}{\bibfnamefont{G.~D.}
  \bibnamefont{{Starkman}}}, \bibinfo{journal}{\prd}
  \textbf{\bibinfo{volume}{75}}, \bibinfo{pages}{023507}
  (\bibinfo{year}{2007}), \eprint{arXiv:astro-ph/0605135}.

\bibitem[{\citenamefont{{de Oliveira-Costa} et~al.}(2004)\citenamefont{{de
  Oliveira-Costa}, {Tegmark}, {Zaldarriaga}, and {Hamilton}}}]{dOCTZH}
\bibinfo{author}{\bibfnamefont{A.}~\bibnamefont{{de Oliveira-Costa}}},
  \bibinfo{author}{\bibfnamefont{M.}~\bibnamefont{{Tegmark}}},
  \bibinfo{author}{\bibfnamefont{M.}~\bibnamefont{{Zaldarriaga}}},
  \bibnamefont{and}
  \bibinfo{author}{\bibfnamefont{A.}~\bibnamefont{{Hamilton}}},
  \bibinfo{journal}{\prd} \textbf{\bibinfo{volume}{69}},
  \bibinfo{pages}{063516} (\bibinfo{year}{2004}),
  \eprint{arXiv:astro-ph/0307282}.

\bibitem[{\citenamefont{{Schwarz} et~al.}(2004)\citenamefont{{Schwarz},
  {Starkman}, {Huterer}, and {Copi}}}]{schwarz}
\bibinfo{author}{\bibfnamefont{D.~J.} \bibnamefont{{Schwarz}}},
  \bibinfo{author}{\bibfnamefont{G.~D.} \bibnamefont{{Starkman}}},
  \bibinfo{author}{\bibfnamefont{D.}~\bibnamefont{{Huterer}}},
  \bibnamefont{and} \bibinfo{author}{\bibfnamefont{C.~J.}
  \bibnamefont{{Copi}}}, \bibinfo{journal}{Physical Review Letters}
  \textbf{\bibinfo{volume}{93}}, \bibinfo{pages}{221301}
  (\bibinfo{year}{2004}), \eprint{arXiv:astro-ph/0403353}.

\bibitem[{\citenamefont{{Copi} et~al.}(2004)\citenamefont{{Copi}, {Huterer},
  and {Starkman}}}]{copi1}
\bibinfo{author}{\bibfnamefont{C.~J.} \bibnamefont{{Copi}}},
  \bibinfo{author}{\bibfnamefont{D.}~\bibnamefont{{Huterer}}},
  \bibnamefont{and} \bibinfo{author}{\bibfnamefont{G.~D.}
  \bibnamefont{{Starkman}}}, \bibinfo{journal}{\prd}
  \textbf{\bibinfo{volume}{70}}, \bibinfo{pages}{043515}
  (\bibinfo{year}{2004}), \eprint{arXiv:astro-ph/0310511}.

\bibitem[{\citenamefont{{Hajian}}(2007)}]{hajian}
\bibinfo{author}{\bibfnamefont{A.}~\bibnamefont{{Hajian}}},
  \bibinfo{journal}{ArXiv Astrophysics e-prints}  (\bibinfo{year}{2007}),
  \eprint{astro-ph/0702723}.

\bibitem[{\citenamefont{{Eriksen} et~al.}(2004)\citenamefont{{Eriksen},
  {Hansen}, {Banday}, {G{\'o}rski}, and {Lilje}}}]{eriksen2004}
\bibinfo{author}{\bibfnamefont{H.~K.} \bibnamefont{{Eriksen}}},
  \bibinfo{author}{\bibfnamefont{F.~K.} \bibnamefont{{Hansen}}},
  \bibinfo{author}{\bibfnamefont{A.~J.} \bibnamefont{{Banday}}},
  \bibinfo{author}{\bibfnamefont{K.~M.} \bibnamefont{{G{\'o}rski}}},
  \bibnamefont{and} \bibinfo{author}{\bibfnamefont{P.~B.}
  \bibnamefont{{Lilje}}}, \bibinfo{journal}{\apj}
  \textbf{\bibinfo{volume}{605}}, \bibinfo{pages}{14} (\bibinfo{year}{2004}).

\bibitem[{\citenamefont{{Freeman} et~al.}(2006)\citenamefont{{Freeman},
  {Genovese}, {Miller}, {Nichol}, and {Wasserman}}}]{freeman}
\bibinfo{author}{\bibfnamefont{P.~E.} \bibnamefont{{Freeman}}},
  \bibinfo{author}{\bibfnamefont{C.~R.} \bibnamefont{{Genovese}}},
  \bibinfo{author}{\bibfnamefont{C.~J.} \bibnamefont{{Miller}}},
  \bibinfo{author}{\bibfnamefont{R.~C.} \bibnamefont{{Nichol}}},
  \bibnamefont{and}
  \bibinfo{author}{\bibfnamefont{L.}~\bibnamefont{{Wasserman}}},
  \bibinfo{journal}{\apj} \textbf{\bibinfo{volume}{638}}, \bibinfo{pages}{1}
  (\bibinfo{year}{2006}), \eprint{arXiv:astro-ph/0510406}.

\bibitem[{\citenamefont{{Campanelli} et~al.}(2006)\citenamefont{{Campanelli},
  {Cea}, and {Tedesco}}}]{ellipsoid}
\bibinfo{author}{\bibfnamefont{L.}~\bibnamefont{{Campanelli}}},
  \bibinfo{author}{\bibfnamefont{P.}~\bibnamefont{{Cea}}}, \bibnamefont{and}
  \bibinfo{author}{\bibfnamefont{L.}~\bibnamefont{{Tedesco}}},
  \bibinfo{journal}{Physical Review Letters} \textbf{\bibinfo{volume}{97}},
  \bibinfo{pages}{131302} (\bibinfo{year}{2006}).

\bibitem[{\citenamefont{{Barrow} et~al.}(1997)\citenamefont{{Barrow},
  {Ferreira}, and {Silk}}}]{barrow}
\bibinfo{author}{\bibfnamefont{J.~D.} \bibnamefont{{Barrow}}},
  \bibinfo{author}{\bibfnamefont{P.~G.} \bibnamefont{{Ferreira}}},
  \bibnamefont{and} \bibinfo{author}{\bibfnamefont{J.}~\bibnamefont{{Silk}}},
  \bibinfo{journal}{Physical Review Letters} \textbf{\bibinfo{volume}{78}},
  \bibinfo{pages}{3610} (\bibinfo{year}{1997}),
  \eprint{arXiv:astro-ph/9701063}.

\bibitem[{\citenamefont{{Ghosh} et~al.}(2007)\citenamefont{{Ghosh}, {Hajian},
  and {Souradeep}}}]{ghosh}
\bibinfo{author}{\bibfnamefont{T.}~\bibnamefont{{Ghosh}}},
  \bibinfo{author}{\bibfnamefont{A.}~\bibnamefont{{Hajian}}}, \bibnamefont{and}
  \bibinfo{author}{\bibfnamefont{T.}~\bibnamefont{{Souradeep}}},
  \bibinfo{journal}{\prd} \textbf{\bibinfo{volume}{75}},
  \bibinfo{pages}{083007} (\bibinfo{year}{2007}),
  \eprint{arXiv:astro-ph/0604279}.

\bibitem[{\citenamefont{{Jaffe} et~al.}(2006)\citenamefont{{Jaffe}, {Banday},
  {Eriksen}, {G{\'o}rski}, and {Hansen}}}]{bianchi}
\bibinfo{author}{\bibfnamefont{T.~R.} \bibnamefont{{Jaffe}}},
  \bibinfo{author}{\bibfnamefont{A.~J.} \bibnamefont{{Banday}}},
  \bibinfo{author}{\bibfnamefont{H.~K.} \bibnamefont{{Eriksen}}},
  \bibinfo{author}{\bibfnamefont{K.~M.} \bibnamefont{{G{\'o}rski}}},
  \bibnamefont{and} \bibinfo{author}{\bibfnamefont{F.~K.}
  \bibnamefont{{Hansen}}}, \bibinfo{journal}{\aap}
  \textbf{\bibinfo{volume}{460}}, \bibinfo{pages}{393} (\bibinfo{year}{2006}),
  \eprint{arXiv:astro-ph/0606046}.

\bibitem[{\citenamefont{{Inoue} and {Silk}}(2006)}]{inoue}
\bibinfo{author}{\bibfnamefont{K.~T.} \bibnamefont{{Inoue}}} \bibnamefont{and}
  \bibinfo{author}{\bibfnamefont{J.}~\bibnamefont{{Silk}}},
  \bibinfo{journal}{\apj} \textbf{\bibinfo{volume}{648}}, \bibinfo{pages}{23}
  (\bibinfo{year}{2006}), \eprint{arXiv:astro-ph/0602478}.

\bibitem[{\citenamefont{{Inoue} and {Silk}}(2007)}]{inoue2}
\bibinfo{author}{\bibfnamefont{K.~T.} \bibnamefont{{Inoue}}} \bibnamefont{and}
  \bibinfo{author}{\bibfnamefont{J.}~\bibnamefont{{Silk}}},
  \bibinfo{journal}{\apj} \textbf{\bibinfo{volume}{664}}, \bibinfo{pages}{650}
  (\bibinfo{year}{2007}), \eprint{arXiv:astro-ph/0612347}.

\bibitem[{\citenamefont{{Abramo} et~al.}(2006)\citenamefont{{Abramo},
  {Sodr{\'e}}, and {Wuensche}}}]{abramo}
\bibinfo{author}{\bibfnamefont{L.~R.} \bibnamefont{{Abramo}}},
  \bibinfo{author}{\bibfnamefont{L.~J.} \bibnamefont{{Sodr{\'e}}}},
  \bibnamefont{and} \bibinfo{author}{\bibfnamefont{C.~A.}
  \bibnamefont{{Wuensche}}}, \bibinfo{journal}{\prd}
  \textbf{\bibinfo{volume}{74}}, \bibinfo{pages}{083515}
  (\bibinfo{year}{2006}), \eprint{arXiv:astro-ph/0605269}.

\bibitem[{\citenamefont{{Cooray} and {Seto}}(2005)}]{cooray}
\bibinfo{author}{\bibfnamefont{A.}~\bibnamefont{{Cooray}}} \bibnamefont{and}
  \bibinfo{author}{\bibfnamefont{N.}~\bibnamefont{{Seto}}},
  \bibinfo{journal}{Journal of Cosmology and Astro-Particle Physics}
  \textbf{\bibinfo{volume}{12}}, \bibinfo{pages}{4} (\bibinfo{year}{2005}),
  \eprint{arXiv:astro-ph/0510137}.

\bibitem[{\citenamefont{{Frisch}}(2005)}]{frisch}
\bibinfo{author}{\bibfnamefont{P.~C.} \bibnamefont{{Frisch}}},
  \bibinfo{journal}{\apjl} \textbf{\bibinfo{volume}{632}},
  \bibinfo{pages}{L143} (\bibinfo{year}{2005}),
  \eprint{arXiv:astro-ph/0506293}.

\bibitem[{\citenamefont{{Efstathiou}}(2004)}]{efstathiou2}
\bibinfo{author}{\bibfnamefont{G.}~\bibnamefont{{Efstathiou}}},
  \bibinfo{journal}{\mnras} \textbf{\bibinfo{volume}{348}},
  \bibinfo{pages}{885} (\bibinfo{year}{2004}), \eprint{arXiv:astro-ph/0310207}.

\bibitem[{\citenamefont{{Szapudi} et~al.}(2001)\citenamefont{{Szapudi},
  {Prunet}, and {Colombi}}}]{spice}
\bibinfo{author}{\bibfnamefont{I.}~\bibnamefont{{Szapudi}}},
  \bibinfo{author}{\bibfnamefont{S.}~\bibnamefont{{Prunet}}}, \bibnamefont{and}
  \bibinfo{author}{\bibfnamefont{S.}~\bibnamefont{{Colombi}}},
  \bibinfo{journal}{\apjl} \textbf{\bibinfo{volume}{561}}, \bibinfo{pages}{L11}
  (\bibinfo{year}{2001}).

\bibitem[{\citenamefont{{Gruppuso}}(2007)}]{ellipsoidiswrong}
\bibinfo{author}{\bibfnamefont{A.}~\bibnamefont{{Gruppuso}}},
  \bibinfo{journal}{\prd} \textbf{\bibinfo{volume}{76}},
  \bibinfo{pages}{083010} (\bibinfo{year}{2007}), \eprint{arXiv:0705.2536}.

\bibitem[{\citenamefont{{Gould}}(1993)}]{gouldquad}
\bibinfo{author}{\bibfnamefont{A.}~\bibnamefont{{Gould}}},
  \bibinfo{journal}{\apjl} \textbf{\bibinfo{volume}{403}}, \bibinfo{pages}{L51}
  (\bibinfo{year}{1993}).

\bibitem[{\citenamefont{{Smoot} et~al.}(1992)\citenamefont{{Smoot}, {Bennett},
  {Kogut}, {Wright}, {Aymon}, {Boggess}, {Cheng}, {de Amici}, {Gulkis},
  {Hauser} et~al.}}]{cobe}
\bibinfo{author}{\bibfnamefont{G.~F.} \bibnamefont{{Smoot}}},
  \bibinfo{author}{\bibfnamefont{C.~L.} \bibnamefont{{Bennett}}},
  \bibinfo{author}{\bibfnamefont{A.}~\bibnamefont{{Kogut}}},
  \bibinfo{author}{\bibfnamefont{E.~L.} \bibnamefont{{Wright}}},
  \bibinfo{author}{\bibfnamefont{J.}~\bibnamefont{{Aymon}}},
  \bibinfo{author}{\bibfnamefont{N.~W.} \bibnamefont{{Boggess}}},
  \bibinfo{author}{\bibfnamefont{E.~S.} \bibnamefont{{Cheng}}},
  \bibinfo{author}{\bibfnamefont{G.}~\bibnamefont{{de Amici}}},
  \bibinfo{author}{\bibfnamefont{S.}~\bibnamefont{{Gulkis}}},
  \bibinfo{author}{\bibfnamefont{M.~G.} \bibnamefont{{Hauser}}},
  \bibnamefont{et~al.}, \bibinfo{journal}{\apjl}
  \textbf{\bibinfo{volume}{396}}, \bibinfo{pages}{L1} (\bibinfo{year}{1992}).

\bibitem[{\citenamefont{{G{\'o}rski} et~al.}(2005)\citenamefont{{G{\'o}rski},
  {Hivon}, {Banday}, {Wandelt}, {Hansen}, {Reinecke}, and
  {Bartelmann}}}]{healpix}
\bibinfo{author}{\bibfnamefont{K.~M.} \bibnamefont{{G{\'o}rski}}},
  \bibinfo{author}{\bibfnamefont{E.}~\bibnamefont{{Hivon}}},
  \bibinfo{author}{\bibfnamefont{A.~J.} \bibnamefont{{Banday}}},
  \bibinfo{author}{\bibfnamefont{B.~D.} \bibnamefont{{Wandelt}}},
  \bibinfo{author}{\bibfnamefont{F.~K.} \bibnamefont{{Hansen}}},
  \bibinfo{author}{\bibfnamefont{M.}~\bibnamefont{{Reinecke}}},
  \bibnamefont{and}
  \bibinfo{author}{\bibfnamefont{M.}~\bibnamefont{{Bartelmann}}},
  \bibinfo{journal}{\apj} \textbf{\bibinfo{volume}{622}}, \bibinfo{pages}{759}
  (\bibinfo{year}{2005}), \eprint{arXiv:astro-ph/0409513}.

\bibitem[{\citenamefont{{Spergel} et~al.}(2007)\citenamefont{{Spergel}, {Bean},
  {Dor{\'e}}, {Nolta}, {Bennett}, {Dunkley}, {Hinshaw}, {Jarosik}, {Komatsu},
  {Page} et~al.}}]{wmap2}
\bibinfo{author}{\bibfnamefont{D.~N.} \bibnamefont{{Spergel}}},
  \bibinfo{author}{\bibfnamefont{R.}~\bibnamefont{{Bean}}},
  \bibinfo{author}{\bibfnamefont{O.}~\bibnamefont{{Dor{\'e}}}},
  \bibinfo{author}{\bibfnamefont{M.~R.} \bibnamefont{{Nolta}}},
  \bibinfo{author}{\bibfnamefont{C.~L.} \bibnamefont{{Bennett}}},
  \bibinfo{author}{\bibfnamefont{J.}~\bibnamefont{{Dunkley}}},
  \bibinfo{author}{\bibfnamefont{G.}~\bibnamefont{{Hinshaw}}},
  \bibinfo{author}{\bibfnamefont{N.}~\bibnamefont{{Jarosik}}},
  \bibinfo{author}{\bibfnamefont{E.}~\bibnamefont{{Komatsu}}},
  \bibinfo{author}{\bibfnamefont{L.}~\bibnamefont{{Page}}},
  \bibnamefont{et~al.}, \bibinfo{journal}{\apjs}
  \textbf{\bibinfo{volume}{170}}, \bibinfo{pages}{377} (\bibinfo{year}{2007}),
  \eprint{arXiv:astro-ph/0603449}.

\bibitem[{\citenamefont{{Spergel} et~al.}(2003)\citenamefont{{Spergel},
  {Verde}, {Peiris}, {Komatsu}, {Nolta}, {Bennett}, {Halpern}, {Hinshaw},
  {Jarosik}, {Kogut} et~al.}}]{wmap1yrparams}
\bibinfo{author}{\bibfnamefont{D.~N.} \bibnamefont{{Spergel}}},
  \bibinfo{author}{\bibfnamefont{L.}~\bibnamefont{{Verde}}},
  \bibinfo{author}{\bibfnamefont{H.~V.} \bibnamefont{{Peiris}}},
  \bibinfo{author}{\bibfnamefont{E.}~\bibnamefont{{Komatsu}}},
  \bibinfo{author}{\bibfnamefont{M.~R.} \bibnamefont{{Nolta}}},
  \bibinfo{author}{\bibfnamefont{C.~L.} \bibnamefont{{Bennett}}},
  \bibinfo{author}{\bibfnamefont{M.}~\bibnamefont{{Halpern}}},
  \bibinfo{author}{\bibfnamefont{G.}~\bibnamefont{{Hinshaw}}},
  \bibinfo{author}{\bibfnamefont{N.}~\bibnamefont{{Jarosik}}},
  \bibinfo{author}{\bibfnamefont{A.}~\bibnamefont{{Kogut}}},
  \bibnamefont{et~al.}, \bibinfo{journal}{\apjs}
  \textbf{\bibinfo{volume}{148}}, \bibinfo{pages}{175} (\bibinfo{year}{2003}),
  \eprint{arXiv:astro-ph/0302209}.

\bibitem[{\citenamefont{{Efstathiou}}(2003)}]{efstathiou}
\bibinfo{author}{\bibfnamefont{G.}~\bibnamefont{{Efstathiou}}},
  \bibinfo{journal}{\mnras} \textbf{\bibinfo{volume}{346}},
  \bibinfo{pages}{L26} (\bibinfo{year}{2003}), \eprint{arXiv:astro-ph/0306431}.

\bibitem[{\citenamefont{{Gazta{\~n}aga}
  et~al.}(2003)\citenamefont{{Gazta{\~n}aga}, {Wagg}, {Multam{\"a}ki},
  {Monta{\~n}a}, and {Hughes}}}]{gaztanaga}
\bibinfo{author}{\bibfnamefont{E.}~\bibnamefont{{Gazta{\~n}aga}}},
  \bibinfo{author}{\bibfnamefont{J.}~\bibnamefont{{Wagg}}},
  \bibinfo{author}{\bibfnamefont{T.}~\bibnamefont{{Multam{\"a}ki}}},
  \bibinfo{author}{\bibfnamefont{A.}~\bibnamefont{{Monta{\~n}a}}},
  \bibnamefont{and} \bibinfo{author}{\bibfnamefont{D.~H.}
  \bibnamefont{{Hughes}}}, \bibinfo{journal}{\mnras}
  \textbf{\bibinfo{volume}{346}}, \bibinfo{pages}{47} (\bibinfo{year}{2003}),
  \eprint{arXiv:astro-ph/0304178}.

\bibitem[{\citenamefont{{Cline} et~al.}(2003)\citenamefont{{Cline}, {Crotty},
  and {Lesgourgues}}}]{cline}
\bibinfo{author}{\bibfnamefont{J.~M.} \bibnamefont{{Cline}}},
  \bibinfo{author}{\bibfnamefont{P.}~\bibnamefont{{Crotty}}}, \bibnamefont{and}
  \bibinfo{author}{\bibfnamefont{J.}~\bibnamefont{{Lesgourgues}}},
  \bibinfo{journal}{Journal of Cosmology and Astro-Particle Physics}
  \textbf{\bibinfo{volume}{9}}, \bibinfo{pages}{10} (\bibinfo{year}{2003}),
  \eprint{arXiv:astro-ph/0304558}.

\bibitem[{\citenamefont{{Contaldi} et~al.}(2003)\citenamefont{{Contaldi},
  {Peloso}, {Kofman}, and {Linde}}}]{contaldi}
\bibinfo{author}{\bibfnamefont{C.~R.} \bibnamefont{{Contaldi}}},
  \bibinfo{author}{\bibfnamefont{M.}~\bibnamefont{{Peloso}}},
  \bibinfo{author}{\bibfnamefont{L.}~\bibnamefont{{Kofman}}}, \bibnamefont{and}
  \bibinfo{author}{\bibfnamefont{A.}~\bibnamefont{{Linde}}},
  \bibinfo{journal}{Journal of Cosmology and Astro-Particle Physics}
  \textbf{\bibinfo{volume}{7}}, \bibinfo{pages}{2} (\bibinfo{year}{2003}),
  \eprint{arXiv:astro-ph/0303636}.

\bibitem[{\citenamefont{{Starobinskij}}(1993)}]{starobinsky}
\bibinfo{author}{\bibfnamefont{A.~A.} \bibnamefont{{Starobinskij}}},
  \bibinfo{journal}{Soviet Journal of Experimental and Theoretical Physics
  Letters} \textbf{\bibinfo{volume}{57}}, \bibinfo{pages}{622}
  (\bibinfo{year}{1993}), \eprint{arXiv:gr-qc/9305019}.

\bibitem[{\citenamefont{{Stevens} et~al.}(1993)\citenamefont{{Stevens},
  {Scott}, and {Silk}}}]{stevens}
\bibinfo{author}{\bibfnamefont{D.}~\bibnamefont{{Stevens}}},
  \bibinfo{author}{\bibfnamefont{D.}~\bibnamefont{{Scott}}}, \bibnamefont{and}
  \bibinfo{author}{\bibfnamefont{J.}~\bibnamefont{{Silk}}},
  \bibinfo{journal}{Physical Review Letters} \textbf{\bibinfo{volume}{71}},
  \bibinfo{pages}{20} (\bibinfo{year}{1993}).

\bibitem[{\citenamefont{{de Oliveira-Costa} et~al.}(1996)\citenamefont{{de
  Oliveira-Costa}, {Smoot}, and {Starobinsky}}}]{angelica}
\bibinfo{author}{\bibfnamefont{A.}~\bibnamefont{{de Oliveira-Costa}}},
  \bibinfo{author}{\bibfnamefont{G.~F.} \bibnamefont{{Smoot}}},
  \bibnamefont{and} \bibinfo{author}{\bibfnamefont{A.~A.}
  \bibnamefont{{Starobinsky}}}, \bibinfo{journal}{\apj}
  \textbf{\bibinfo{volume}{468}}, \bibinfo{pages}{457} (\bibinfo{year}{1996}),
  \eprint{arXiv:astro-ph/9510109}.

\bibitem[{\citenamefont{{Bunn} and {Scott}}(2000)}]{bonehead}
\bibinfo{author}{\bibfnamefont{E.~F.} \bibnamefont{{Bunn}}} \bibnamefont{and}
  \bibinfo{author}{\bibfnamefont{D.}~\bibnamefont{{Scott}}},
  \bibinfo{journal}{\mnras} \textbf{\bibinfo{volume}{313}},
  \bibinfo{pages}{331} (\bibinfo{year}{2000}), \eprint{arXiv:astro-ph/9906044}.

\bibitem[{\citenamefont{{Luminet} et~al.}(2003)\citenamefont{{Luminet},
  {Weeks}, {Riazuelo}, {Lehoucq}, and {Uzan}}}]{luminet}
\bibinfo{author}{\bibfnamefont{J.-P.} \bibnamefont{{Luminet}}},
  \bibinfo{author}{\bibfnamefont{J.~R.} \bibnamefont{{Weeks}}},
  \bibinfo{author}{\bibfnamefont{A.}~\bibnamefont{{Riazuelo}}},
  \bibinfo{author}{\bibfnamefont{R.}~\bibnamefont{{Lehoucq}}},
  \bibnamefont{and} \bibinfo{author}{\bibfnamefont{J.-P.}
  \bibnamefont{{Uzan}}}, \bibinfo{journal}{\nat}
  \textbf{\bibinfo{volume}{425}}, \bibinfo{pages}{593} (\bibinfo{year}{2003}),
  \eprint{arXiv:astro-ph/0310253}.

\bibitem[{\citenamefont{{Cornish} et~al.}(2004)\citenamefont{{Cornish},
  {Spergel}, {Starkman}, and {Komatsu}}}]{circles}
\bibinfo{author}{\bibfnamefont{N.~J.} \bibnamefont{{Cornish}}},
  \bibinfo{author}{\bibfnamefont{D.~N.} \bibnamefont{{Spergel}}},
  \bibinfo{author}{\bibfnamefont{G.~D.} \bibnamefont{{Starkman}}},
  \bibnamefont{and}
  \bibinfo{author}{\bibfnamefont{E.}~\bibnamefont{{Komatsu}}},
  \bibinfo{journal}{Physical Review Letters} \textbf{\bibinfo{volume}{92}},
  \bibinfo{pages}{201302} (\bibinfo{year}{2004}),
  \eprint{arXiv:astro-ph/0310233}.

\bibitem[{\citenamefont{{Gordon} et~al.}(2005)\citenamefont{{Gordon}, {Hu},
  {Huterer}, and {Crawford}}}]{gordon}
\bibinfo{author}{\bibfnamefont{C.}~\bibnamefont{{Gordon}}},
  \bibinfo{author}{\bibfnamefont{W.}~\bibnamefont{{Hu}}},
  \bibinfo{author}{\bibfnamefont{D.}~\bibnamefont{{Huterer}}},
  \bibnamefont{and}
  \bibinfo{author}{\bibfnamefont{T.}~\bibnamefont{{Crawford}}},
  \bibinfo{journal}{\prd} \textbf{\bibinfo{volume}{72}},
  \bibinfo{pages}{103002} (\bibinfo{year}{2005}),
  \eprint{arXiv:astro-ph/0509301}.

\end{thebibliography}

\end{document}